# Calculation of structural symmetry and applicability of small $A_kB_lC_m$ nanoclusters (A=Rb, Cs; B=Pb, Sn, Bi; C=Cl, Br, I) as building blocks for synthesis of larger-size nanoclusters

*Makarevych R.,*

*Dmytruk I., Hrabovskyi Ye., Pundyk I., Bashmakova N., Berezovska N.*

*Taras Shevchenko National University of Kyiv*

## ABSTRACT

In this report we analyze charasteristics of structural symmetry of small $A_kB_lC_m$ nanoclusters (A=Rb, Cs; B=Pb, Sn, Bi; C=Cl, Br, I), binding energies, stability, and, ultimately, applicability as building blocks for construction of larger-size nanoclusters [1].

Preliminary spatial modeling of ternary compound nanoclusters has been carried out on the basics of geometrical approach, namely interatomic distances taken as a sums of corresponding ionic radii. The geometry optimization and binding energies of the modeled structures have been determined through quantum-chemical calculations. Geometry optimization has been performed with Gaussian09 software package using Density Functional Theory (DFT) within the generalized gradient approximation of the Perdew-Burke-Ernzerhof (PBE) and hybrid approximation of the Becke-3-Lee-Yang-Parr (B3LYP) exchange-correlation functionals. The Stuttgart/Dresden (SDD) and Los Alamos National Laboratory 2 Double Zeta (LANL2DZ) scalar-relativistic effective basic potentials were employed throughout the calculations.

As a result of these calculations, a clear relationship between the atomic radii, their ratios within the nanoclusters and their stability has been obtained and analyzed.

## 1. Introduction

Nanoparticles of so-called perovskite semiconductors consisting of a heavy alkali metal, a post-transition metal (mostly Pb), and a halogen have drawn much interest in the last decades due to their unique luminescent and optoelectronic properties and promising applications. However, most of the studies deal with quantum dots of perovskites – nanoparticles in several nanometer size range that preserve stoichiometry and structure of corresponding bulk crystals.

Nanoclusters (sub-2 nm nanoparticles) is the class of materials that bridges the gap between molecules and quantum dots nanoparticles and are characterized by multiple unique properties. Some of these properties are similar to quantum dots arising from quantum confinement and discrete, molecule-like electronic structure, such as discrete energy levels, increased bandgap and blue shift of absorption/emission spectra. This makes them highly promising for various practical applications such as bioimaging, chemical catalysis, and data storage among others. But nanoclusters have also unique properties, for example, structural symmetry never observed for larger nanoparticles.

Magic-size nanoclusters demonstrate elevated stability among others in the certain mass range and consequently are favorable for applications. One of such applications is the use of them as chemical building blocks for larger nanoclusters [1]. Although multiple methods of large (10-100nm) nanoclusters have been extensively described in scientific literature ([2], [3], [4]), seed-mediated growth method based on magic-sized nanoclusters is still rather unexplored.

Small-size $A_kB_lC_m$ nanoclusters (A=Rb, Cs; B=Pb, Sn, Bi; C=Cl, Br, I) have been extensively described in academic literature as nanomaterials with a number of unique properties and, correspondingly, a vast array of practical applications [3].

We perform an overview of the possibility of such application of small nanoclusters based on their sizes, structural symmetry, atomic radii and their ratios. The core criteria for nanocluster viability as a chemical building block is its stability, which can be precisely numerically determined by binding energy per atom. This parameter can be calculated using the formula:

$$E_b = \frac{[E_{tot}(A_kB_lC_m) - kE_{tot}(A) - lE_{tot}(B) - mE_{tot}(C)]}{k+l+m},$$

where $E_b$ – binding energy per atom, $E_{tot}$ – absolute energy of cluster/molecule/atom, *k, l, m* – quantities of atoms of elements A, B and C respectively.

To determine correlations of nanoclusters stability with their various characteristics we conducted comparative analysis of 6 selected nanoclusters ($Cs_3PbI_5$, $Rb_3SnBr_5$, $Cs_2BiI_5$, $Rb_2BiBr_5$, $Cs_4SnBr_4$, $Rb_4SnI_4$) grouped in pairs by geometry. All of the nanoclusters selected for the analysis were modeled to have a core-shell stereometric structure, where the shell was formed by alkaline metal (Rb, Cs) and halogen (Br, I) atoms and the core was represented by singular transition metal (Sn, Pb, Bi) atom. The choice of this structure and elements is substantiated by the property of heavy-element 0D perovskites to show record-high energy conversion efficiency stemming from highly discreet zone structure [8]. The specific characteristics which we considered in context of their influence on nanoclusters' stability were average atomic radius, shell size (relative to a similar core) and atomic radius ratio within a shell.

For comparison of binding energies per atom for each nanocluster we created their starting idealized geometry models using the Chem3D software package, performed geometry optimization and subsequent absolute energy calculation via density functional theory (DFT) method with PBE/SDD and B3LYP/LANL2DZ exchange-correlation functional/effective core basic potential pairs using GAUSSIAN16 software. The density functional theory method was chosen for these quantum chemical calculations for manageability of its computational weight and history of its successful usage for similar calculations in scientific literature [1]. The former feature arises from DFT's nature as a computational QM modeling method that determines electronic properties of many-body systems (primarily in ground state) using functionals of spatially dependent electronic density which forms a bijective relationship with wave function of analyzed system:

$$n(r) = N \int \partial^3 r_2 \dots \int \partial^3 r_N \Psi^*(r, r_2, \dots, r_N)\Psi(r, r_2, \dots, r_N),$$

where $r,\dots r_N$ - electron coordinates, $\Psi$ - system's wave function, $N$ – quantity of electrons in the system.

This provides a way to systematically map many-body problem with electron-electron interaction potential onto a single-body problem without the electron-electron interaction potential which allows to avoid solving many-body Schrödinger equation directly thus significantly reducing computational weight. In addition, most common potential weaknesses of DFT such as describing intermolecular interactions, charge transfer excitations, transition states and some strongly correlated systems [9], have little to no influence on quality of results of performed calculations.

The accuracy of a DFT calculation depends strongly on the approximation used for the exchange–correlation contribution to the total energy. In the present work, the generalized-gradient approximation represented by PBE and the hybrid approximation represented by B3LYP were used. These approaches differ in their treatment of exchange and correlation since PBE, like all generalized-gradient approximations, interprets exchange-correlation energy as function of electron density and its gradient while B3LYP, as hybrid functional, combines Hartree-Fock exact exchange functional, generalized gradient approximation (the Becke 88 exchange functional [5] and the correlation functional of Lee, Yang and Parr [6]) and localized spin density approximation to the correlation functional [7]. This difference results in B3LYP being potentially more accurate but also more computationally heavy. To offset this difference and make results directly comparable, B3LYP was used with computationally lighter LANL2DZ basis set, which only utilizes double-zeta valence functions (pairs of 2 distinct basis functions with different exponents (zeta values)) while PBE was used with SDD, which acts as a higher quality basis set for heavy elements often utilizing triple-zeta valence functions for them (thus also increasing its inherent computational weight).

Both approaches include relativistic effective core potentials (ECP) for inner shell electrons and therefore can provide a practical balance between computational efficiency and the ability to describe the electronic structure of relatively complex heavy-element-containing nanoclusters. The presence of Rb, Cs, Sn, Pb, Bi, Br, and I additionally makes the use of effective core potentials particularly relevant because these atoms contain large numbers of inner-shell electrons. The SDD and LanL2DZ approaches therefore provide computationally tractable descriptions of such systems while effectively incorporating important core-related relativistic effects through the corresponding ECP treatment. The use of two independent functional–basis/ECP combinations made it possible not only to compare the stability of individual nanoclusters but also to examine whether the observed relationships between atomic radii and energetic stability are robust with respect to the computational approach.

## 2. Results and discussion

The investigated systems were arranged into three pairs, each designed to isolate a particular size-related factor. The first pair, $Cs_3PbI_5$ and $Rb_3SnBr_5$, was used to investigate the relationship between the average atomic radius of the cluster and its stability. The second pair, $Cs_2BiI_5$ and $Rb_2BiBr_5$, was selected to analyze the influence of shell size while maintaining a similar central atomic species. Finally, the pair $Cs_4SnBr_4 > Rb_4SnI_4$ was considered to study the influence of the ratio between the atomic radii of the components forming the outer shell of the nanocluster. An important common feature of these comparisons is that the selected pairs retain similar structural symmetry after optimization.

Consequently, the observed differences in binding energy can primarily be discussed in terms of atomic-size relations and their influence on the optimized interatomic geometry rather than as a direct consequence of a transition between substantially different symmetry classes.

**Effect of the average atomic radius: Cs3PbI5 and Rb3SnBr5**

The first comparison concerns the relationship between the average atomic radius of the nanocluster and its energetic stability. The calculated binding energies per atom for $Rb_3SnBr_5$ and $Cs_3PbI_5$ are summarized by the two computational approaches. Using PBE/SDD, the binding energy per atom was found to be 7.73 Hartree for $Cs_3PbI_5$ and 8.41 Hartree for $Rb_3SnBr_5$. The corresponding B3LYP/LanL2DZ values were 7.19 and 7.77 Hartree, respectively.

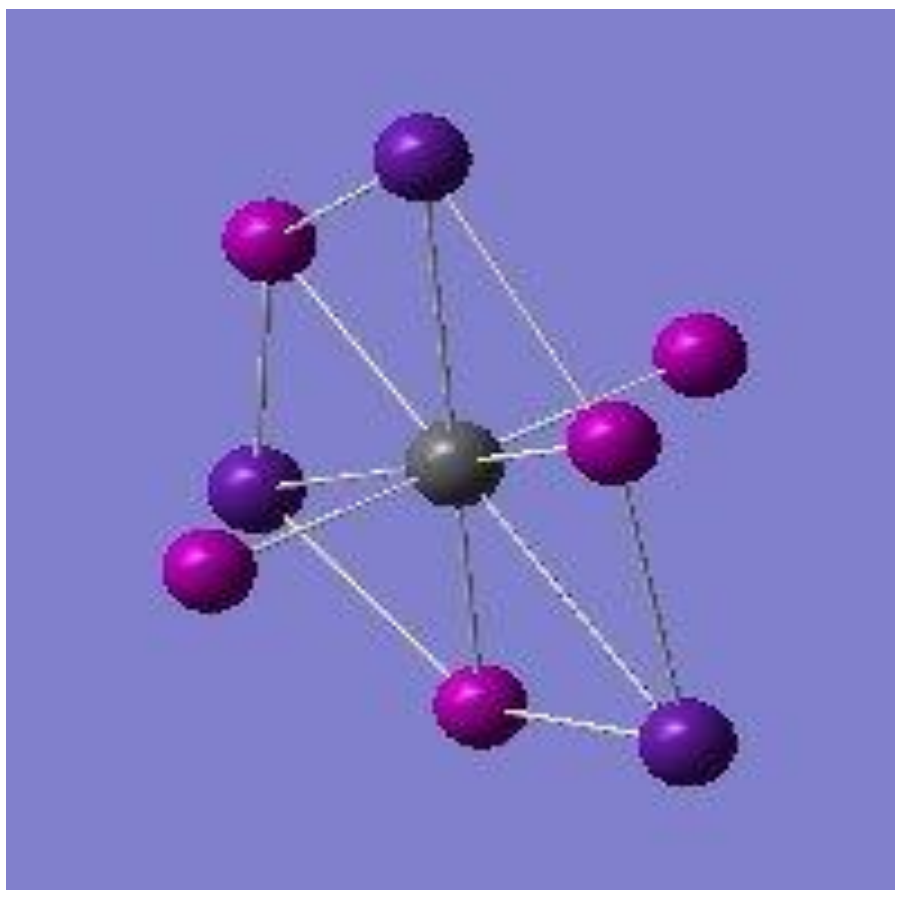

Pic.1 Generalized geometry of $Cs_3PbI_5/Rb_3SnBr_5$ nanoclusters

Thus, both computational approaches predict the same energetic ordering within this pair:

$Cs_3PbI_5 < Rb_3SnBr_5$

with respect to binding energy per atom. This agreement is particularly important because the absolute numerical values obtained by PBE/SDD and B3LYP/LanL2DZ differ, whereas the qualitative stability trend remains unchanged. The result therefore does not depend solely on the specific exchange–correlation functional or ECP combination employed.

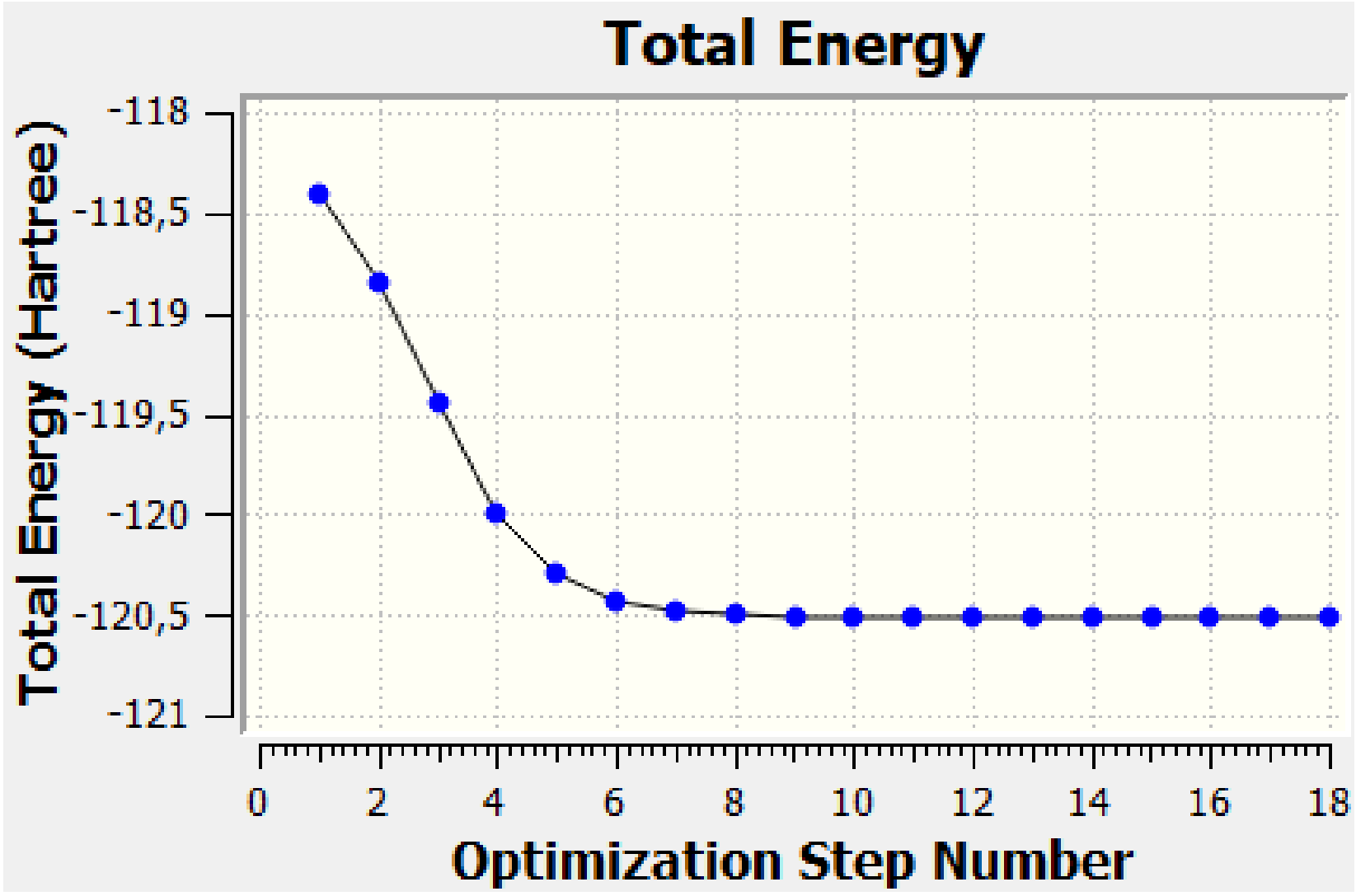

Pic.2 Stepwise geometry optimization curve for $Cs_3PbI_5/Rb_3SnBr_5$ nanoclusters

The comparison indicates a decrease in stability with increasing average atomic radius, i.e. the nanocluster $Rb_3SnBr_5$ , which represents the smaller average atomic-size scale within the selected comparison, possesses a higher binding energy per atom than $Cs_3PbI_5$. From a structural point of view, this behavior can be interpreted as reflecting the influence of atomic dimensions on the characteristic interatomic separations and on the degree of spatial compactness that can be achieved after geometry optimization. When the constituent atoms become larger, the equilibrium geometry generally expands, which can modify the balance between attractive and repulsive interactions and reduce the average energetic stabilization per atom.

At the same time, the results should not be interpreted as implying that the average atomic radius is an independent universal predictor of nanocluster stability. The chemical identities of the atoms also change between the two clusters. Therefore, the average radius should be considered here as a structural descriptor representing the overall size scale of the system rather than as the only microscopic variable controlling the energy. Nevertheless, the fact that both DFT approaches reproduce the same trend demonstrates that the average atomic size is a useful parameter for comparative analysis of structurally related clusters.

An additional observation is that the symmetry of the optimized structures remains unchanged within the comparison considered in the presentation. This suggests that the change in stability occurs without a substantial symmetry transformation. Consequently, the energetic effect can be associated with continuous changes in atomic dimensions and equilibrium distances rather than with the appearance or disappearance of symmetry elements. This distinction is important for the general objective of identifying potential chemical building blocks: two nanoclusters can possess comparable symmetry properties while exhibiting noticeably different energetic stability.

**Effect of shell size with a similar core: Cs2BiI5 and Rb2BiBr5**

The second pair was selected to investigate the influence of the outer-shell dimensions while keeping the central atomic component similar. The calculated PBE/SDD binding energies per atom were 7.23 Hartree for $Cs_2BiI_5$ and 8.26 Hartree for $Rb_2BiBr_5$. The B3LYP/LanL2DZ calculations produced values of 6.23 and 7.29 Hartree, respectively.

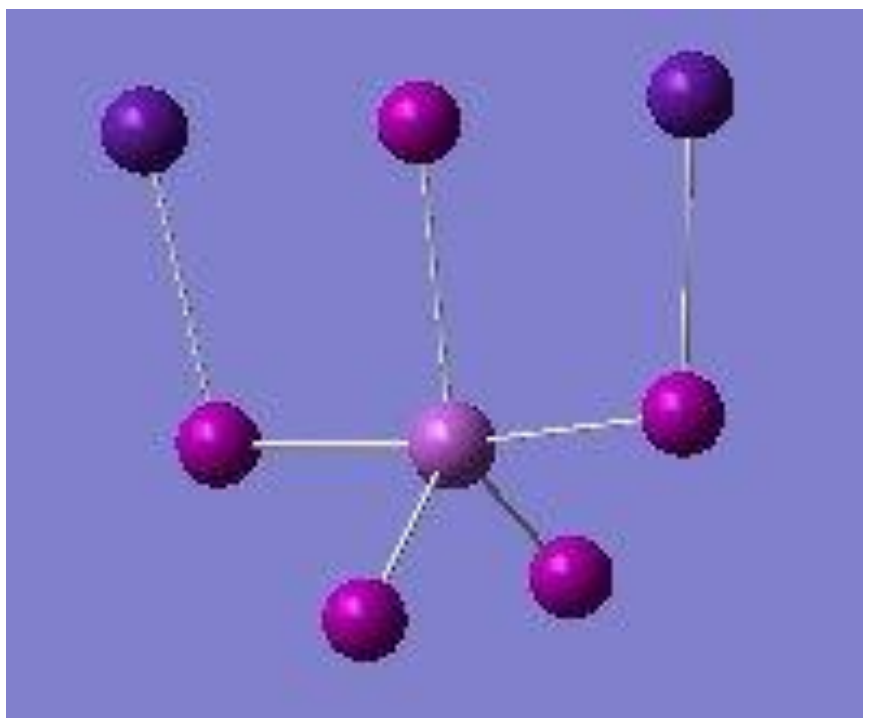

Pic. 3 Generalized geometry of $Cs_2BiI_5/Rb_2BiBr_5$ nanoclusters

Again, both computational methods provide the same qualitative ordering:

$Cs_2BiI_5 < Rb_2BiBr_5$

This result demonstrates a consistent decrease of binding energy per atom with increasing shell size for the selected pair, i.e. replacement of the smaller Rb/Br shell environment by the larger Cs/I environment increases the characteristic spatial dimensions of the nanocluster. The resulting optimized structure is therefore expected to have a larger overall geometrical scale and longer characteristic interatomic distances.

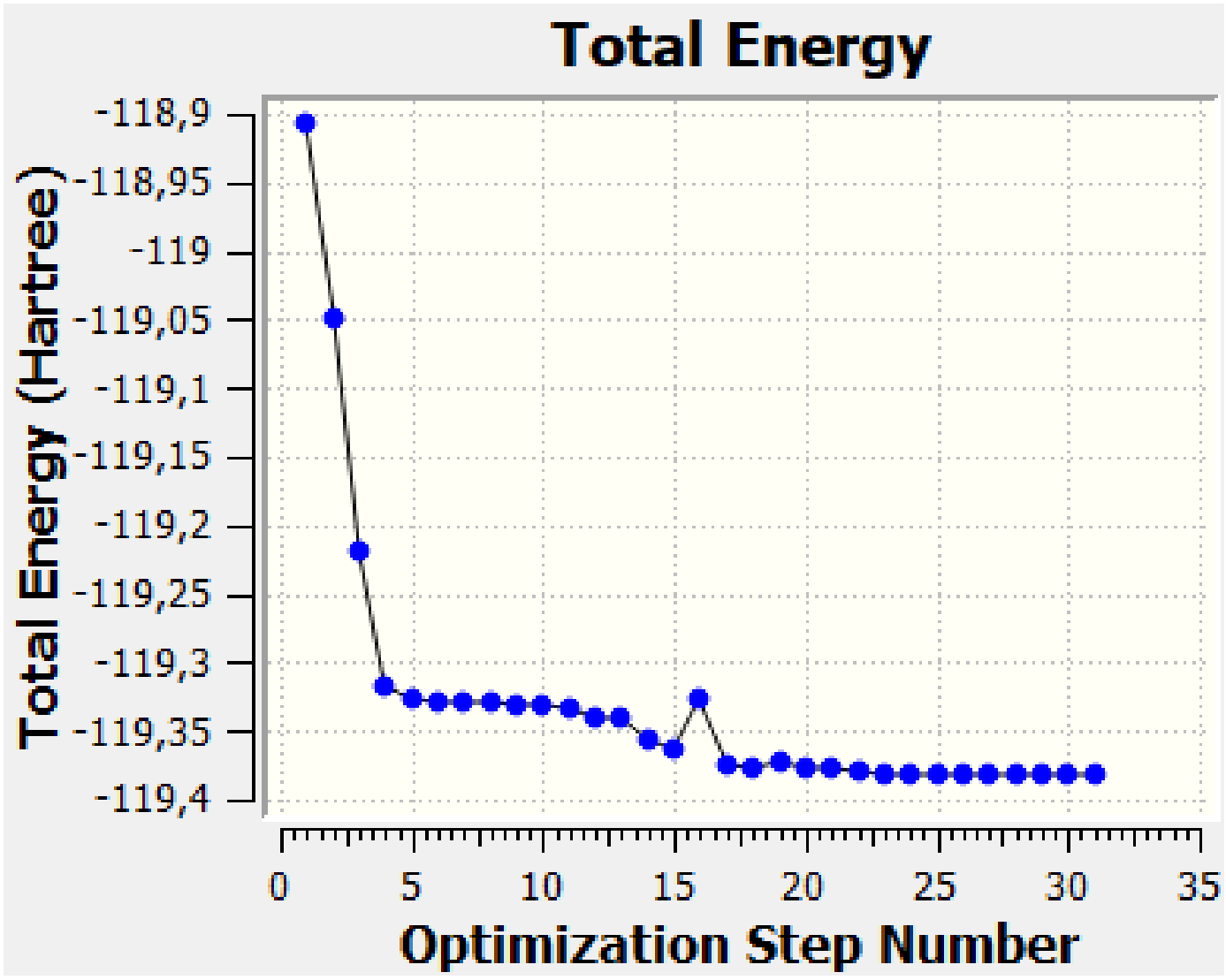

Pic.4 Stepwise geometry optimization curve for $Cs_2BiI_5/Rb_2BiBr_5$ nanoclusters

The importance of this comparison lies in the possibility of separating, at least qualitatively, the effect of the central atom from the effect of the surrounding shell. Since the Bi-containing core remains common to both systems, the observed change in stability can be associated primarily with the modification of the atoms forming the shell. This indicates that, for nanoclusters of this structural type, a more compact shell can provide a more favorable average binding environment.

The reduction in stability associated with an increase in shell size may be understood in terms of geometrical compactness and the redistribution of interatomic interactions throughout the cluster. A larger shell requires a larger spatial arrangement around the central region. Such expansion can decrease the average strength of the interactions contributing to the stabilization of the cluster when the energy is normalized per atom. Importantly, the conclusion concerns relative stability within the present family of related structures and should not be generalized to all nanocluster systems without additional calculations.

As in the first comparison, the symmetry after optimization remains unchanged. This is a significant observation because it demonstrates that stability can vary substantially even when the principal symmetry properties of the structures are retained. Therefore, structural symmetry by itself cannot be considered a sufficient energetic criterion for selecting a nanocluster as a chemical building block. Instead, symmetry should be treated as one structural descriptor among several, while the binding energy per atom provides an independent quantitative measure of energetic viability.

The agreement between PBE/SDD and B3LYP/LanL2DZ is especially clear for this pair. Although B3LYP/LanL2DZ produces lower absolute values of the binding energy per atom, both approaches predict the same stability ordering and the same direction of the size effect. Such agreement strengthens confidence in the conclusion that the observed tendency is not an accidental consequence of one particular computational protocol.

**Effect of the shell atomic-radius ratio: Cs4SnBr4 and Rb4SnI4**

The third comparison examines a different structural parameter: the ratio between the atomic radii of the components forming the shell of the nanocluster. The calculated PBE/SDD binding energies per atom were 7.74 Hartree for $Cs_4SnBr_4$ and 7.02 Hartree for $Rb_4SnI_4$. The corresponding B3LYP/LanL2DZ values were 7.16 and 6.51 Hartree.

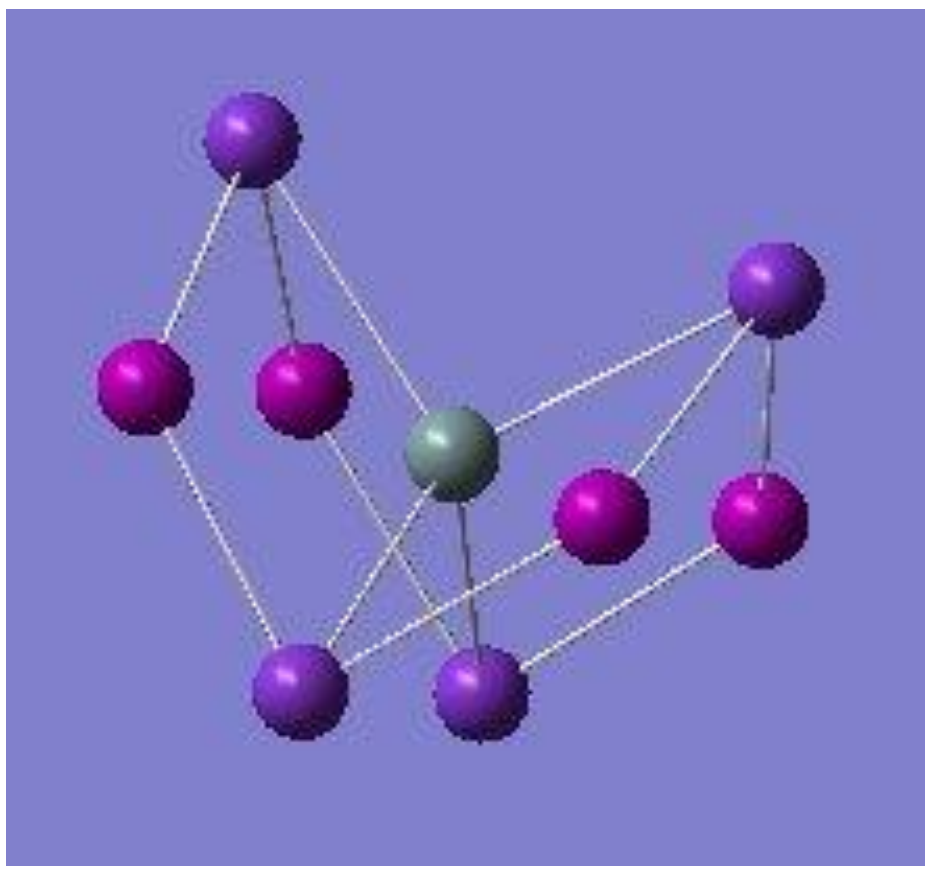

Pic.5 Generalized geometry of $Cs_4SnBr_4/Rb_4SnI_4$ nanoclusters

Both methods therefore predict:

$Cs_4SnBr_4 > Rb_4SnI_4$

According to the comparative analysis presented for these structures, the stability increases with an increase in the shell atomic-radius ratio. This result differs conceptually from the first two comparisons. The average size of the cluster or the absolute size of the shell is not the principal descriptor in this case. Instead, the relative size matching between the atoms occupying different positions in the shell is considered.

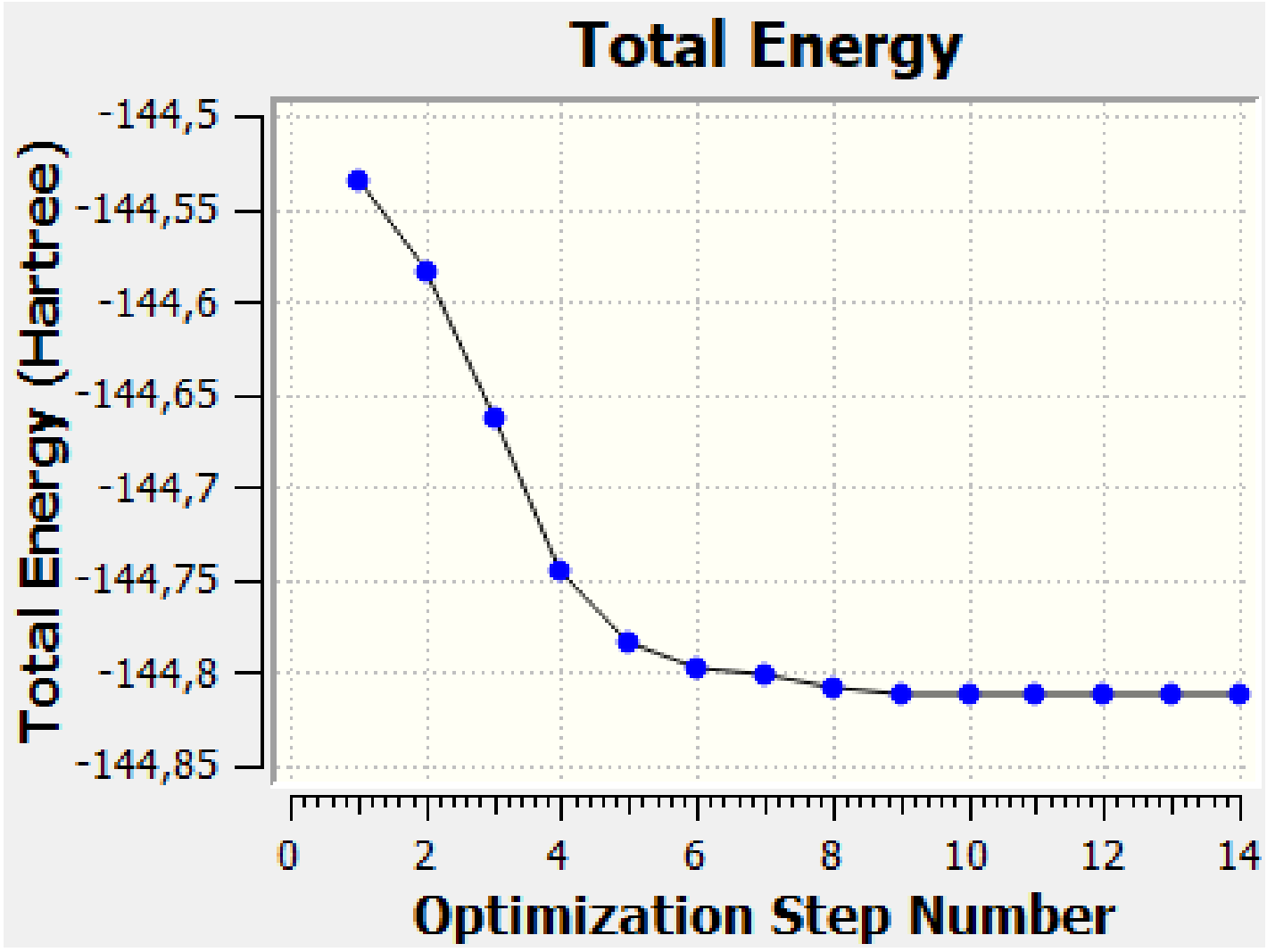

Pic.6 Stepwise geometry optimization curve for $Cs_4SnBr_4/Rb_4SnI_4$ nanoclusters

This observation emphasizes that atomic-size effects in nanoclusters cannot be reduced to a simple rule stating that smaller atoms always produce more stable structures. The relative dimensions of the interacting atoms are also important. A particular ratio of atomic radii can provide a more favorable geometrical fit between the different components of the structure. Such size matching can influence the optimized bond lengths and the distribution of strain within the cluster, thereby changing the total energetic stabilization.

The $Cs_4SnBr_4/Rb_4SnI_4$ comparison therefore provides an important complement to the other two pairs. While the first pair shows a negative correlation between average atomic radius and stability, and the second pair demonstrates a negative correlation between shell size and stability, the third pair shows that an appropriate relative size relationship between shell atoms can increase stability. Taken together, these observations indicate that the energetics of small $A_kB_lC_m$ nanoclusters are controlled by several interconnected geometrical parameters.

Once again, the symmetry remains unchanged within the pair. The difference in stability is consequently not associated with a major change of the symmetry class but with changes in the dimensions and relative arrangement of the atoms constituting the same general structural framework. This provides an important result for the proposed application of small nanoclusters as chemical building blocks. Structural symmetry may be preserved over a range of compositions, but the energetic suitability of the corresponding building blocks can still differ significantly because of atomic-size matching.

**Comparative assessment of PBE/SDD and B3LYP/LanL2DZ**

The complete set of calculated binding energies provides a basis for comparing the two computational approaches. For all six considered nanoclusters, PBE/SDD gives higher absolute binding-energy values than B3LYP/LanL2DZ. However, the most important result for the present comparative study is not the systematic difference in absolute values but the degree to which both approaches reproduce the same qualitative trends.

| **Eb, Hartree $*10^{-2}$** | **$Cs_3PbI_5$** | **$Rb_3SnBr_5$** | **$Cs_2BiI_5$** | **$Rb_2BiBr_5$** | **$Cs_4SnBr_4$** | **$Rb_4SnI_4$** |
|---|---|---|---|---|---|---|
| PBE/SDD | 7.73 | 8.41 | 7.23 | 8.26 | 7.74 | 7.02 |
| B3LYP/LANL2DZ | 7.19 | 7.77 | 6.23 | 7.29 | 7.16 | 6.51 |

Tab.1 Binding energies per atom for all considered nanoclusters

For the three directly investigated pairs, both methods predict identical conclusions. Both show a decrease in stability with increasing average atomic radius for the $Cs_3PbI_5/Rb_3SnBr_5$ comparison. Both show a decrease in stability with increasing shell size for the $Cs_2BiI_5/Rb_2BiBr_5$ comparison. Finally, both show an increase in stability with the shell atomic-radius ratio for the $Cs_4SnBr_4/Rb_4SnI_4$ pair.

The general stability ranking also shows substantial agreement. Both approaches identify $Rb_3SnBr_5$ as the most stable of the considered structures and $Rb_2BiBr_5$ as the second most stable. Differences become more noticeable for the clusters with closer binding-energy values. In particular, the relative ordering of $Cs_3PbI_5$ and $Cs_4SnBr_4$, as well as that of $Cs_2BiI_5$ and $Rb_4SnI_4$, is sensitive to the computational approach. This indicates that the two methods are highly consistent for identifying pronounced stability differences but may provide different rankings when the energetic differences between clusters are relatively small.

This observation leads to an important methodological conclusion. PBE/SDD and B3LYP/LanL2DZ can both be considered useful tools for comparative analysis of nanoclusters with similar structural and symmetry properties. The consistent reproduction of the principal trends demonstrates that both approaches are capable of providing meaningful data for identifying size–stability relationships. At the same time, differences in the detailed ranking of energetically similar clusters show that absolute stability predictions should be interpreted with caution.

For systematic screening of larger families of related nanoclusters, PBE/SDD has a practical advantage because it combines a computationally efficient GGA functional with an effective core potential particularly suitable for systems containing heavy atoms. B3LYP/LanL2DZ provides a useful independent comparison level and can serve as a validation method for representative structures or for cases in which the stability ranking predicted by PBE/SDD is uncertain. The two methods should therefore be regarded as complementary rather than mutually exclusive.

The results also define the limits of the present comparison. The reliability demonstrated here applies primarily to nanoclusters with similar structural symmetry and comparable geometrical frameworks. If substantially different symmetry classes are compared, changes in binding topology and electronic structure may become sufficiently important that simple size-based trends and direct comparisons of binding energies require additional verification. Consequently, a computational protocol validated for comparative screening within one structural family should not automatically be assumed to provide the same reliability for clusters with strongly different geometries.

Overall, the calculations demonstrate that atomic-size relations provide useful structural descriptors for understanding the stability of small $A_kB_lC_m$ nanoclusters. The average atomic radius, the absolute size of the shell, and the ratio of the radii of shell components each influence the energetic properties of the optimized structures. At the same time, these factors do not necessarily cause noticeable changes in symmetry when the starting idealized geometries belong to similar structural families. This makes it possible to distinguish between the role of symmetry as a structural descriptor and the role of binding energy per atom as a direct measure of energetic stability. Such a distinction is essential for selecting small, stable nanoclusters that may subsequently be considered as chemical building blocks for the construction of larger nanocluster systems.

## 3. Conclusions

The comparative quantum-chemical analysis of six small $A_kB_lC_m$ nanoclusters containing heavy alkali metals, p-block elements, and halogens demonstrates that their energetic stability is strongly influenced by relationships between the sizes of their constituent atoms. The results obtained for the three selected pairs show that atomic radii should not be considered merely as independent elemental characteristics. Instead, their average values, relative ratios, and distribution between the core and shell collectively affect the optimized geometrical structure and, consequently, the binding energy per atom of the nanocluster.

The comparison of $Cs_3PbI_5$ and $Rb_3SnBr_5$ demonstrates that, for the investigated structural pair, stability decreases with increasing average atomic radius. Both PBE/SDD and B3LYP/LanL2DZ predict the same qualitative trend, with $Rb_3SnBr_5$ exhibiting a higher binding energy per atom than $Cs_3PbI_5$. This agreement indicates that the relationship between the overall atomic-size scale of the cluster and its energetic stability is robust with respect to the two computational approaches used in the present work.

The analysis of $Cs_2BiI_5$ and $Rb_2BiBr_5$ further shows that the dimensions of the outer shell constitute another important factor. Both computational methods predict a decrease in stability when the shell size increases. The smaller-shell nanocluster $Rb_2BiBr_5$ is more stable than $Cs_2BiI_5$ according to both PBE/SDD and B3LYP/LanL2DZ calculations. This result supports the conclusion that increased geometrical compactness of the shell can provide a more favorable energetic arrangement for small nanoclusters of similar structural type.

The third comparison, involving $Cs_4SnBr_4$ and $Rb_4SnI_4$, demonstrates that relative atomic dimensions are also important. In this case, both computational approaches show an increase in stability with the shell atomic-radius ratio. Thus, the results indicate that nanocluster stability cannot be described by a simple dependence on absolute atomic size alone. The geometrical compatibility and relative size matching of the atoms occupying different positions in the cluster can significantly influence the energetic stabilization of the optimized structure.

An important general result is that the considered atomic-size variations do not produce noticeable changes in the symmetry properties of the selected nanoclusters, provided that their starting idealized geometries belong to similar structural families. Therefore, the observed changes in binding energy occur predominantly without substantial symmetry transformations. This finding shows that structural symmetry and energetic stability, although potentially related, should not be treated as equivalent characteristics. A nanocluster with preserved or high symmetry is not necessarily the most stable member of a structural family. Symmetry should consequently be used as a structural descriptor, whereas the binding energy per atom provides an independent quantitative criterion for evaluating energetic stability and the potential viability of a nanocluster as a chemical building block.

The comparison of PBE/SDD and B3LYP/LanL2DZ demonstrates substantial consistency between the two computational approaches. Although PBE/SDD systematically produces higher absolute binding-energy values, both methods reproduce the same principal trends for all three selected pairs. They therefore provide consistent conclusions concerning the effects of average atomic radius, shell size, and shell atomic-radius ratio. The general stability ranking is also substantially similar, especially for the most stable structures, although some differences appear for nanoclusters whose calculated binding energies are close to one another.

On this basis, both PBE/SDD and B3LYP/LanL2DZ can be considered reliable computational tools for comparative investigations of nanoclusters with similar symmetry properties and related structural frameworks. For systematic screening of larger families of heavy-element-containing nanoclusters, PBE/SDD represents a particularly practical primary computational protocol because of its relatively favorable computational efficiency and the applicability of the SDD effective core potential to heavy atoms. B3LYP/LanL2DZ can simultaneously be used as an independent validation level for representative structures and for cases in which small differences in stability require additional verification.

However, the applicability of these approaches should not be generalized without qualification to nanoclusters possessing substantially different symmetries or strongly different binding topologies. Under such conditions, changes in electronic structure and geometrical organization may become more significant than the size-related trends identified in the present study. Additional benchmarking and higher-level calculations may therefore be required when comparing nanoclusters belonging to distinctly different structural families.

In summary, the present results support the use of atomic-size-related parameters as effective descriptors in the preliminary screening of small $A_kB_lC_m$ nanoclusters. Decreasing average atomic radius and shell size, as well as favorable shell atomic-radius ratios, can contribute to increased energetic stability within the investigated structural families. The combination of structural symmetry analysis with the calculation of binding energy per atom therefore provides a useful framework for identifying stable nanoclusters that may serve as potential chemical building blocks for the synthesis and design of larger nanocluster systems.

Present research is supported by the NRFU within the project 2025.07/0410 “Structural regularities and physical properties of magic nanoclusters of ternary compounds”.